\documentclass[reprint,amsmath,amssymb, aps,pre,floatfix,superscriptaddress]{revtex4-2}

\usepackage{graphicx} 
\usepackage{siunitx}
\usepackage[version=4]{mhchem}
\usepackage{comment}
\usepackage{xcolor}
\usepackage{tikz}
\usepackage{scalerel}
\usepackage{xifthen}
\usepackage{chngcntr}
\usepackage{xcolor}
\usepackage{bm}
\usepackage{physics}
\usepackage[utf8]{inputenc}
\usepackage[T1]{fontenc}

\usetikzlibrary{svg.path}
\usepackage[colorlinks,allcolors=blue]{hyperref}

\definecolor{orcidlogocol}{HTML}{A6CE39}
\tikzset{
  orcidlogo/.pic={
    \fill[orcidlogocol] svg{M256,128c0,70.7-57.3,128-128,128C57.3,256,0,198.7,0,128C0,57.3,57.3,0,128,0C198.7,0,256,57.3,256,128z};
    \fill[white] svg{M86.3,186.2H70.9V79.1h15.4v48.4V186.2z}
 svg{M108.9,79.1h41.6c39.6,0,57,28.3,57,53.6c0,27.5-21.5,53.6-56.8,53.6h-41.8V79.1z M124.3,172.4h24.5c34.9,0,42.9-26.5,42.9-39.7c0-21.5-13.7-39.7-43.7-39.7h-23.7V172.4z}
 svg{M88.7,56.8c0,5.5-4.5,10.1-10.1,10.1c-5.6,0-10.1-4.6-10.1-10.1c0-5.6,4.5-10.1,10.1-10.1C84.2,46.7,88.7,51.3,88.7,56.8z};
  }
}

\newcommand\orcid[1]{\href{https://orcid.org/#1}{\mbox{\scalerel*{
\begin{tikzpicture}[yscale=-1,transform shape]
\pic{orcidlogo};
\end{tikzpicture}
}{|}}}}

\begin{document}

\setlength{\unitlength}{1cm}

\def \De {\text{De}}
\def \Re {\text{Re}}
\def \Ma {\text{Ma}}
\def \Rt {\mR}

\def \mA {\mathcal{A}}
\def \mD {\mathcal{D}}
\def \mC {\mathcal{C}}
\def \mF {\mathcal{F}}

\def \mI {\mathcal{I}}
\def \mL {\mathcal{L}}
\def \mM {\mathcal{M}}
\def \mN {\mathcal{N}}
\def \mT {\mathcal{T}}
\def \mR {\mathcal{R}}
\def \mU {\mathcal{U}}
\def \mV {\mathcal{V}}
\def \mO {\mathcal{O}}
\def \mX {\mathcal{X}}
\def \mH {\mathcal{H}}
\def \mh {\mathcal{h}}

\def \ba {\mathbf{a}}
\def \bb {\mathbf{b}}
\def \be {\mathbf{e}}
\def \bf {\mathbf{f}}

\def \bbf {\mathbf{f}}
\def \bq {\mathbf{q}}
\def \br {\mathbf{r}}
\def \bt {\mathbf{t}}

\def \bu {\mathbf{u}}
\def \bv {\mathbf{v}}
\def \bw {\mathbf{w}}
\def \bx {\mathbf{x}}
\def \by {\mathbf{y}}

\def \bsigma {\boldsymbol{\sigma}}
\def \btau {\boldsymbol{\tau}}

\def \bzero {\boldsymbol{0}}

\def \bA {\mathbf{A}}
\def \bB {\mathbf{B}}
\def \bC {\mathbf{C}}
\def \bD {\mathbf{D}}
\def \bE {\mathbf{E}}
\def \bF {\mathbf{F}}
\def \bG {\mathbf{G}}
\def \bH {\mathbf{H}}
\def \bI {\mathbf{I}}
\def \bJ {\mathbf{J}}
\def \bK {\mathbf{K}}
\def \bL {\mathbf{L}}
\def \bM {\mathbf{M}}
\def \bN {\mathbf{N}}
\def \bO {\mathbf{O}}
\def \bP {\mathbf{P}}
\def \bQ {\mathbf{Q}}
\def \bR {\mathbf{R}}
\def \bS {\mathbf{S}}
\def \bT {\mathbf{T}}
\def \bU {\mathbf{U}}
\def \bV {\mathbf{V}}
\def \bX {\mathbf{X}}

\def \bOmega {\boldsymbol{\Omega}}
\def \bomega {\boldsymbol{\omega}}
\def \btheta {\boldsymbol{\theta}}
\def \bTheta {\boldsymbol{\Theta}}

\def \bell {\boldsymbol{\ell}}

\def \bGamma {\boldsymbol{\Gamma}}

\def \bn {\mathbf{n}}
\def \bI {\mathbf{I}}
\def \tbb {\tilde{\bb}}
\def \tbx {\tilde{\bx}}
\def \tbu {\tilde{\bu}}
\def \tbr {\tilde{\br}}
\def \tbR {\tilde{\bR}}
\def \tbU {\tilde{\bU}}
\def \tbE {\tilde{\bE}}
\def \tbF {\tilde{\bF}}
\def \tbX {\tilde{\bX}}

\def \tbOmega {\tilde{\bOmega}}
\def \tbGamma {\tilde{\bGamma}}
\def \tbtau {\tilde{\btau}}
\def \tbsigma {\tilde{\bsigma}}

\def \ta {\tilde{a}}
\def \tb {\tilde{b}}
\def \tc {\tilde{c}}
\def \td {\tilde{d}}
\def \th {\tilde{h}}
\def \tH {\tilde{H}}
\def \tp {\tilde{p}}
\def \tt {\tilde{t}}
\def \tu {\tilde{u}}
\def \tx {\tilde{x}}
\def \ty {\tilde{y}}

\def \tF {\tilde{F}}
\def \tG {\tilde{G}}

\def \tU {\tilde{U}}
\def \tV {\tilde{V}}
\def \tX {\tilde{X}}
\def \tGamma {\tilde{\Gamma}}
\def \tOmega {\tilde{\Omega}}
\def \ttheta {\tilde{\theta}}
\def \tgrad {\tilde{\grad}}
\def \tpi {\tilde{\pi}}
\def \tdelta {\tilde{\delta}}
\def \tlambda {\tilde{\lambda}}

\def \nprim {\mathcal{N}_{\text{prm}}}
\def \nmf {\mathcal{N}_{\text{mf}}}
\def \nmq {\mathcal{N}_{\text{mq}}}
\def \nbf {\mathcal{N}_{\text{bf}}}
\def \nbq {\mathcal{N}_{\text{bq}}}

\def \aprim {a_{\text{prm}}}
\def \Aprim {A_{\text{prm}}}

\def \aadj {a_{\text{adj}}}

\def \lp {\left(}
\def \rp {\right)}

\def \ls {\left[}
\def \rs {\right]}

\def \d {\text{d}}
\def \dr {\d r}

\def \tran {\mathsf{T}}

\def \ibX {\bX_{\rm im}}
\def \invivo {\textit{in vivo }}

\def \hsep {h_{\rm sep}}

\newcommand{\nus}{Department of Mechanical Engineering, National University of Singapore, 117575, Singapore}
\def \titlep {Enabling microrobotic chemotaxis via reset-free hierarchical reinforcement learning} 
\def \viz {\textit{viz.},~}
\def \ie {\textit{i.e.},~}
\def \eg {\textit{e.g.},~}
\def \mat {Materials and Methods}
\def \SIe {SI}
\def \surf {\partial_D}
\def \Omegamag {\hat{\Omega}}
\def \thetamag {\hat{\theta}}
\def \Qd {Q_\mathrm{d}}
\def \Cc {C_\mathrm{c}}
\def \Us {U_\mathrm{s}}
\def \Uv {U_\mathrm{v}}
\def \NCc {\nabla C_\mathrm{c}}
\def \bS {\mathbf{S}}
\def \bFr {\bF^\text{rep}}
\def \Fr {F^\text{rep}}
\def \tbFr {\tbF^\text{rep}}
\def \tFr {\tF^\text{rep}}
\def \Uswim {\mU}
\def \dm {d_\text{min}}
\newcommand{\movref}[1]{Movie S{#1}}

\def \la {\langle}
\def \ra {\rangle}
\def \re {r}

\def \reone {\re_{1}}
\def \retwo {\re_{2}}

\def \pf {p_{\text{f}}}
\def \pr {p_{\text{r}}}

\def \tpib {\tilde{\pi}_{\beta}}
\def \Vp {V_{\phi}}
\def \Dt {\Delta t}
\def \dt {\delta t}

\title{\titlep}

\author{Tongzhao Xiong~\orcid{0009-0004-5736-4240}}
\affiliation{\nus}
\author{Zhaorong Liu~\orcid{0000-0002-3158-3902}}
\affiliation{\nus}
\author{Chong Jin  Ong~\orcid{0000-0003-0493-4575}}
\affiliation{\nus}
\author{Lailai Zhu~\orcid{0000-0002-3443-0709}}
\email{lailai\_zhu@nus.edu.sg}
\affiliation{\nus}
\date{\today}

\begin{abstract}
Microorganisms have evolved diverse strategies to propel in viscous fluids, navigate complex environments, and exhibit taxis in response to stimuli. This has inspired the development of synthetic microrobots, where machine learning (ML) is playing an increasingly important role. Can ML endow these robots with intelligence resembling that developed by their natural counterparts over evolutionary timelines? Here, we demonstrate chemotactic navigation of a multi-link articulated microrobot using two-level hierarchical reinforcement learning (RL). The lower-level RL allows the robot---featuring either a chain or ring topology---to acquire topology-specific swimming gaits: wave propagation characteristic of flagella or body oscillation akin to an ameboid. Such flagellar and ameboid microswimmers, further enabled by the higher-level RL, accomplish chemotactic navigation in prototypical biologically-relevant scenarios that feature conflicting chemoattractants, pursuing a swimming bacterial mimic, steering in vortical flows, and squeezing through tight constrictions. Additionally, we achieve reset-free, partially observable RL, where the robot observes only its joint angles and local scalar quantities. This advancement illuminates solutions for overcoming the persistent challenges of manual resets and partial observability in real-world microrobotic RL.
\end{abstract}

\maketitle
Machine learning (ML)~\cite{jordan2015machine} has revolutionized the development of robotics, endowing them with higher levels of adaptivity, autonomy, and intelligence. Specifically, reinforcement learning (RL)~\cite{sutton2018reinforcement}---an ML framework---enables robotics to autonomously adapt their behaviours and discover optimal strategies via persistent trial-and-error interactions with their environment. The success and promising prospects of RL in classical robotics~\cite{kober2013reinforcement} have motivated its application in other emerging robotic systems, such as microrobots. These robots, typically micron-sized, untethered, and remotely controlled, represent a cutting-edge medical technology for localised treatment, precise diagnosis, and targeted therapies~\cite{nelson2010microrobots,li2017micro,palagi2018bioinspired,ceylan2019translational,iacovacci2024medical}. 

Notably, these medical applications require an effective navigation system that can accurately direct microrobots to their intended sites of action, typically located within confined, hard-to-reach, and sensitive regions~\cite{yang2021motion}. Overcoming biological barriers \invivo intensifies this requirement: penetration into gel-like extracellular matrix or other non-Newtonian bodily fluids demands excessive thrust to drive the microrobots~\cite{wu2020medical,spagnolie2023swimming}; navigation in branched blood vessels entails guiding them through geometrically complex domains with unsteady and even turbulent flows; deployment near inflammation sites necessitate evasive maneuvers to dodge immune cell attacks~\cite{yasa2020elucidating}.
These complexities collectively make the control environments for microrobots nonlinear, unpredictable, and even chaotic, which are highly difficult to model accurately (if feasible at all). 
This differs from the more deterministic, easier-to-model scenarios typical of traditional robotic systems, such as robotic manipulators, which often support model-based control strategies. Contrastingly, the hard-to-model microrobotic environment precludes these strategies, therefore
strengthening the necessity of incorporating RL and ML in general into microrobotic navigation.

\begin{figure*}[hbt!]
\centering
\includegraphics[width=1\linewidth]{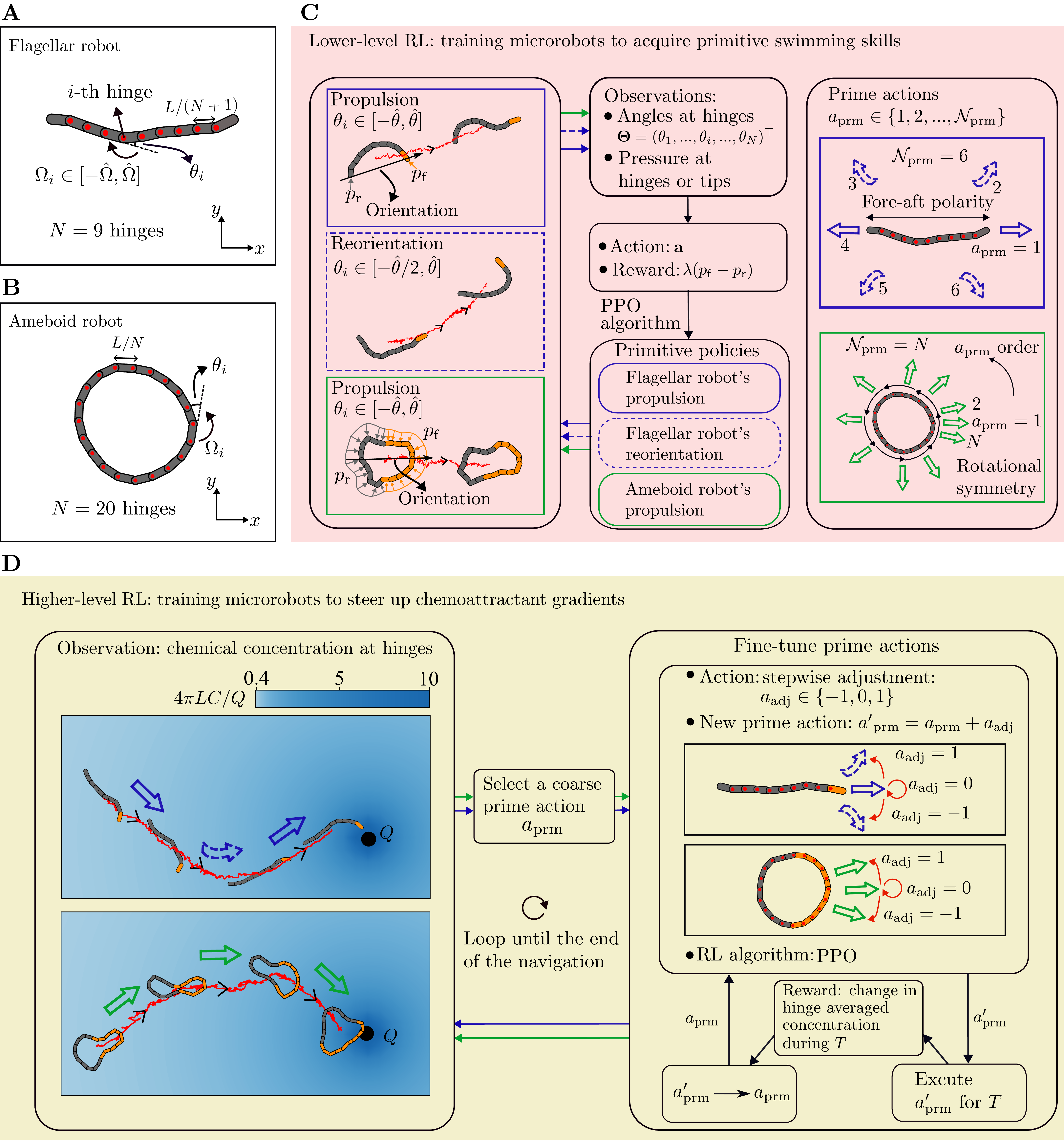}
\caption{
\textbf{Two-level hierarchical RL framework for chemotactic navigation of multi-link microrobots in viscous fluids.} The robot adopts either a linear (A) or ring (B) topology. C, workflow of the lower-level RL: training microrobots to acquire primitive swimming skills---propulsion and reorientation. D, workflow of the higher-level RL: training microrobots to steer up chemoattractant gradients.
}
\label{fig1:RL}
\end{figure*}

RL-based control of microrobots, and more broadly, microswimmers, is an emerging field~\cite{tsang2020roads,cichos2020machine,nasiri2023optimal,yang2024machine,jiao2024deep}. Most studies have been limited to simulated environments, which involve strokes of swimmers~\cite{tsang2020self,qin2023reinforcement,abdi2023self,lin2024emergence,xu2024training}, their navigation in laminar~\cite{schneider2019optimal,gunnarson2021learning, zou2022gait,nasiri2022reinforcement,monderkamp2022active,qiu2022navigation,mo2023chemotaxis,el2023steering,sankaewtong2023learning,putzke2023optimal} and turbulent flows~\cite{biferale2019zermelo,alageshan2020machine},
pursuit~\cite{borra2022reinforcement,zhu2022optimizing,nasiri2024smart}, evasion~\cite{borra2022reinforcement}, and cloaking~\cite{mirzakhanloo2020active}.
In these numerical studies, RL agents are trained in an episodic task, segmented into multiple episodes each featuring a starting and ending point. This episodic training requires resetting the environment to  specific initial states at the beginning of each episode. 
Although straightforward in simulated environments, as adopted in the aforementioned studies, such resets pose significant challenges in experimental settings, representing a major bottleneck in real-world robotic RL~\cite{zhu2020ingredients,ibarz2021train}. The challenge of resetting is exacerbated by the small, confined viscous fluid environments typical of microrobots, thereby resulting in scarce experimental demonstrations of microrobotic RL~\cite{muinos2021reinforcement,behrens2022smart,abbasi2024autonomous}.

Among these experiments, a $5\times 5$ grid environment adopted by the pioneering investigation~\cite{muinos2021reinforcement} eased manual resets, yet it remains idealized compared to real-world microrobotic scenarios. Alternatively, the need for such resets was bypassed by restricting the robotic motion to a circular track~\cite{behrens2022smart}, elegantly leveraging its unique rotational symmetry, which, however, lacks generalizability. Recent experiments~\cite{abbasi2024autonomous} employing more general and realistic configurations have exploited simulations to diminish the frequency of manual resets, although they have not been completely eliminated.

\begin{figure*}[bt!]
\centering
\includegraphics[width=1.0\linewidth]{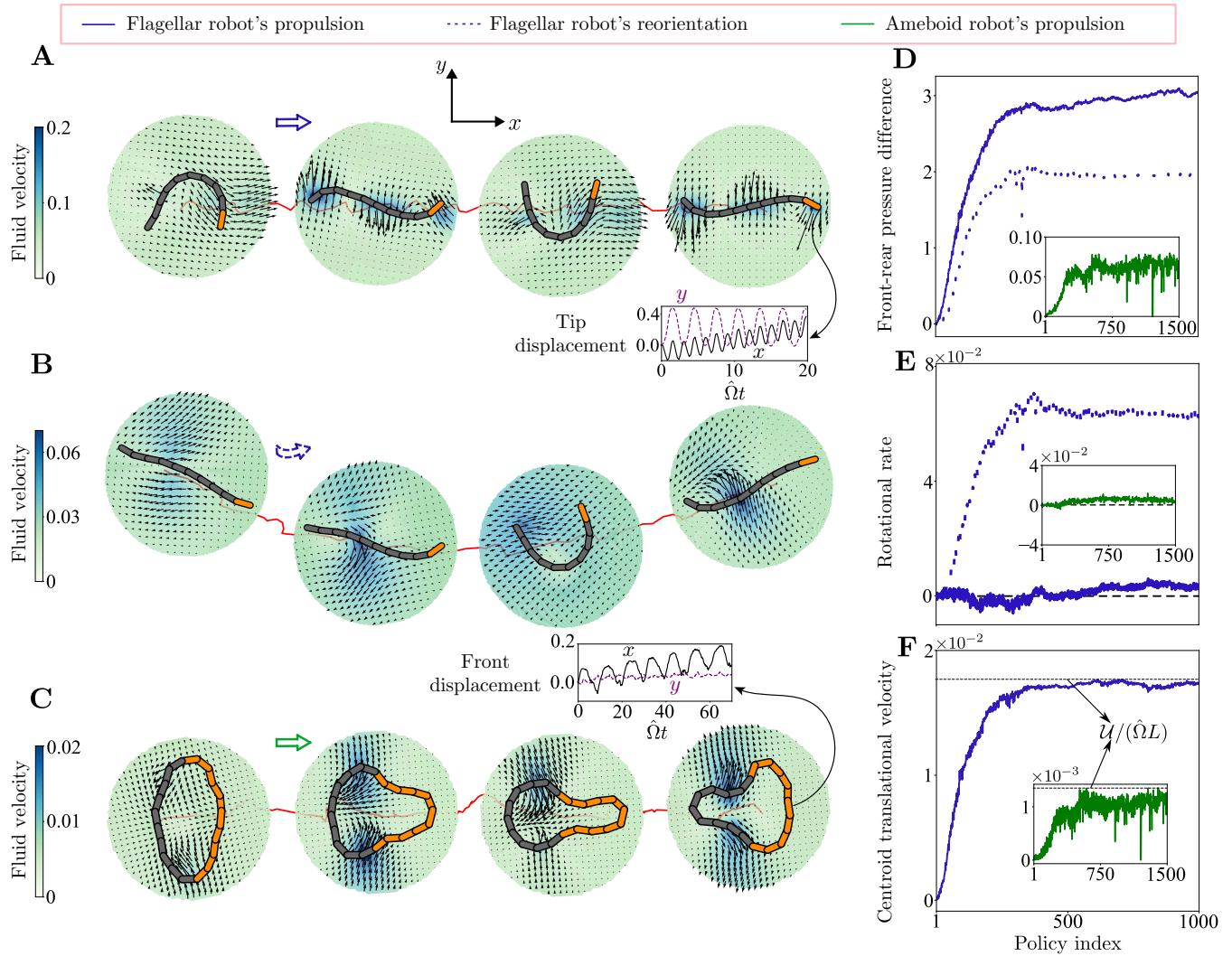}
\caption{
\textbf{Learning primitive swimming skills: propulsion and reorientation.} A, a flagellar robot achieves propulsion through symmetric beating, with vectors indicating the flow field. The inset shows the horizontal ($x$) and vertical ($y$) displacements of the robot's tip, resembling a travelling transverse wave. B, the flagellar robot reorients through asymmetric actuation. C, an ameboid robot propels by periodically contracting and expanding its body. The inset displays the $x$ and $y$ displacements of the robot's front, typical of a longitudinal wave propagating horizontally. 
D, evolution of front-rear pressure differences during the policy iteration process: training the flagellar robot for propulsion (purple solid line) and reorientation (purple dashed), and the ameboid robot for propulsion (green solid). This quantity is time-averaged over one policy. E and F, similar to D, but for the rotational rate and centroid translational velocity of the swimmers, respectively. The latter reaches a plateau value (dashed line), $\Uswim/( \Omegamag L )$, representing the swimming speed of well-trained robots. 
}\label{fig2:policy}
\end{figure*}

\begin{figure*}[t]
\centering
\includegraphics[width=0.9\linewidth]{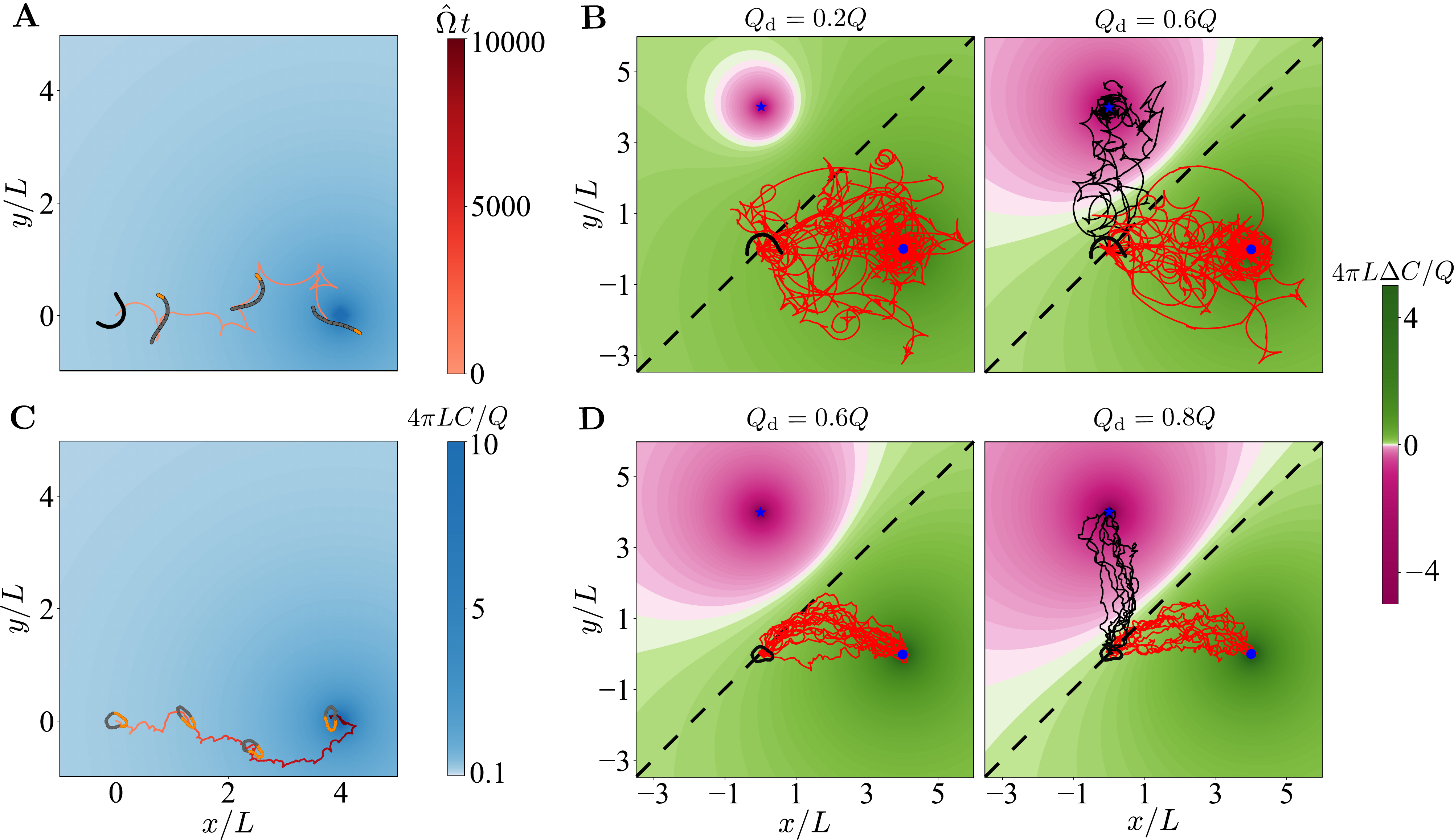}
\caption{\textbf{Chemotaxis towards a stationary chemical source with and without chemical disturbances.} A, a flagellar robot initially at $(0,0)$ swims towards the target source at $(4,0)$. B, similar to A, but with a disturbing chemical source (star symbol) at $(0,4)$ featuring strengths of $\Qd=0.2 Q$ (left panel) and $0.6Q$ (right panel), respectively. Here, $\Delta C$ signifies the difference in the chemoattractant concentration produced by the target and disturbance sources. C, akin to A, but for an ameboid robot. D, analogous to B, but for an ameboid swimmer, with the disturbing source strengths of $\Qd=0.6 Q$ (left panel) and $0.8Q$ (right panel), respectively. 
}
\label{fig:nobg}
\end{figure*}

Besides the cumbersome nature of microrobotic resets, previous studies have generally overlooked observability issues in experiments.
Typically, the agents are assumed to instantaneously obtain robotic velocities, positions, and/or orientations in the global frame, or the near-body global flow velocity~\cite{el2023steering}, or local fluid strain rate and the robot's rotation relative to the local flow vorticity~\cite{qiu2022navigation}. Notably, with the current state-of-the-art technologies, \invivo imaging of a single microrobot presents significant challenges~\cite{nelson2023delivering}, thereby complicating the acquisition of global vectorial signals. On the other hand, local fluid velocities can be mechanosensed by plankton to aid their navigation~\cite{martens2015size}. However, biologically sensing such vectorial/tensorial signals (velocity and strain) typically involves detecting deformation of cellular structures---flagella/cilia and membrane, which are difficult to engineer for microrobots.

In this work,  we draw inspiration from cellular chemotaxis to propose a reset-free hierarchical RL framework that enables the propulsion and chemotactic navigation of microrobots without accessing global or vectorial observations. Our inspiration comes specifically from spermatozoa navigating up a gradient of a chemoattractant (\eg progesterone) released by the egg~\cite{eisenbach2006sperm}, and from neutrophils swimming like an ameboid~\cite{barry2010dictyostelium} toward inflamed sites or pursuing invasive pathogens. 
Accordingly, we design an articulated microrobot consisting of multiple links in two topologies---linear and ring (Fig.~\ref{fig1:RL}A and B): the links are either arranged in a segmented fashion  or connected end-to-end in a loop via a number of $N\geq 3 $ hinges. The former corresponds to a generalized Purcell's swimmer~\cite{purcell1977life,qin2023reinforcement}.

Using a numerical environment, we first demonstrate that the lower level of our hierarchical RL teaches robots with linear and ring topologies to master topology-specific swimming strokes, reminiscent of their biological counterparts. Specifically, the former propels like a whipping flagellum, while the latter deforms in a manner resembling ameboid movement. Utilizing the learnt strokes, the higher level of the hierarchical RL enables both robots to achieve chemotactic navigation under diverse scenarios: steering upwards a static chemical source with and without disturbance cues, pursuing a moving source, and chemotaxing within a background flow or through tight constrictions.

More specifically, both the flagellar and ameboid robots, characterized by a contour length of $L$,
can move in viscous fluids by dynamically rotating their $N$ hinges to adjust the joint angles, $\bTheta=(\theta_1,...,\theta_i,...,\theta_{N})^{\tran}$, where $\theta_i$ corresponds to the $i$-th hinge between two adjacent links (see Fig.~\ref{fig1:RL}A). The angular velocities $\bOmega= \dot{\bTheta}=(\Omega_1,...,\Omega_i,...,\Omega_{N})^{\tran}$ of the hinges represent the swimmer's stroke. Here, $\Omega_i \in [-\Omegamag,\Omegamag]$, with $\Omegamag$ denoting the maximum rotational rate. 
To mimic chemotaxis, the robots sense the local concentration $C$ of a chemoattractant and learn to swim up its gradients. The robots are constrained to perform $xy$-planar locomotion in a three-dimensional (3D) space, with the low-Reynolds-number swimming hydrodynamics approximated by the Stokes equation. Hence, we use a regularized Stokeslet method to emulate the flow environment~\cite{smith2018nearest} (\SIe). 
Further, we assume that chemical diffusion dominates over advection, allowing us to solve the Laplace equation $\grad^2 C = 0$ for $C$.

We consider a non-episodic RL setup consisting of a reset-free learning agent interacting with a partially observable 
environment. The agent lacks access to the robot's position or orientation, rather,  
it relies on proprioceptive sensory inputs, specifically the joint angles $\bTheta$. 
Additionally, the agent receives exteroceptive sensory inputs in the form of hydrodynamic pressure $p$ and chemical concentration $C$ measured at hinges or tips of the multi-link robot.
These inputs, $ \bTheta$, $p$, and $C$ constitute the observations of RL agents.

\subsection*{Hierarchical RL framework for chemotactic navigation}
We develop a two-level hierarchical RL framework to equip the robotic agent with the capacity for autonomous chemotactic navigation (see Fig.~\ref{fig1:RL}). Firstly, the robot acquires two primitive skills, propulsion---moving forward/backward, and reorientation---adjusting swimming directions, in viscous fluids. Secondly, the robot learns to sequence these skills for swimming up the chemical gradients. 
For both levels, we utilize the proximal policy optimization (PPO) algorithm~\cite{schulman2017proximal} implemented in the RLlib library~\cite{liang2018rllib}.

\subsubsection*{Learning primitive swimming skills}\label{sec:primitive}
At this level, the robot learns sequences of a vectorial action $\ba$ of length $N$ to master propulsion and reorientation in an unbounded domain. Here, $\ba$ is linearly associated with the angular velocity $\bOmega$ of hinges, namely the stroke (see \mat).
The flagellar robot masters these two skills through separate RL processes. 
Conversely, the ameboid counterpart requires only the propulsion ability, from which it can derive reorientation. This difference stems from the geometric features of the two robots.
The ameboid robot exhibits a high ($N$) order of rotational symmetry without an intrinsic polarity, allowing it to flexibly maneuver in $N$ directions. This flexibility is achieved by $\nprim=N$ primitive actions $\aprim \in \left\{ 1,2,...,\nprim \right\}$, with each representing a polarity acquired by the robot to propel along (see \mat). Notably, the ameboid swimmer with a large $N$ would resemble a flying saucer.
In contrast, the flagellar swimmer featuring an intrinsic fore-aft polarity cannot execute sideways movement conveniently, thereby requiring reorientation skills to enhance its manoeuvrability. Specifically, these skills enable it to redirect itself in the left-forward, right-forward, left-backward, and right-backward directions with respect to its orientation, in addition to its forward and backward propelling abilities.  Overall, the flagellar swimmer possesses $\nprim=6$ primitive actions.

We design a reward $\reone$ using a locally sensed scalar, pressure $p$, rather than vectorial global quantities such as the swimmer's velocity/displacement as commonly used. Specifically, the reward $\reone = \lambda (\pf-\pr)$ quantifies the pressure differential between the front 
($\pf$) and rear ($\pr$) of the swimmer, measured at the flagellar tips or averaged over the ameboid's halves (Fig.~\ref{fig1:RL}C), see \mat. 
This reward determines the swimming direction: from the rear to the front tip for flagellar swimmers, and from the center of the rear half to that of the front half for ameboid swimmers.

\subsubsection*{Chemotactic navigation utilizing primitive actions}
The higher-level policy aims to identify an optimal sequence of primitive actions within 
to enable efficient chemoattractive motion of the microrobot. The successful realization of this goal is exemplified, in Fig.~\ref{fig1:RL}D, by the movement of microrobots navigating towards a chemical source of strength $Q$.

At this level, we implement a two-step approach: a coarse selection followed by an adjustment. In the coarse step, a set of primitive actions is determined based on the detected chemical concentration at hinges (see \mat). From this set, one action $\aprim$ is randomly selected---not necessarily the optimal one. Subsequently, we tune $\aprim$ by a stepwise adjustment $\aadj$ using RL, 
resulting in a refined action $\aprim^{\prime} = \aprim + \aadj$. Next,  this primitive action $\aprim^{\prime}$ is executed over a fixed interval $T$. Notably, the adjustment action space $\aadj\in\left\{-1,0,1 \right\}$ encompasses a rightward shift ($-1$), no shift  ($0$), and a leftward shift  ($+1$) relative to the coarse action $\aprim$. This adjustment process is repeated across multiple intervals of $T$, with rewards scaling linearly with the cumulative change in hinge-averaged concentration during each interval.

\subsection*{Results}
We focus on a  flagellar robot with $N=9$ hinges and an  ameboid robot with $N=20$, with results for other $N$ presented in \SIe.  We first show the acquisition of primitive swimming skills by these robots. Subsequently, we demonstrate their chemotactic navigation in diverse biologically relevant settings.

\subsubsection*{Topology-specific motility mechanisms}\label{sec:prim_res}
As illustrated in Fig.~\ref{fig2:policy}A (\movref 1), the flagellar swimmer learns to propel by actuating the hinges cyclically, thus causing its body to bend like a 
travelling transverse wave, a strategy harnessed by biological flagella~\cite{taylor1951analysis}. Likewise, the ameboid swimmer acquires propulsion through periodic contraction and expansion, reminiscent of a longitudinal wave, see Fig.~\ref{fig2:policy}C (\movref 3).
This suggests that RL endows topologically distinct robots with topology-specific locomotion mechanisms, effectively aligning them with their biological counterparts. For propulsion, the joint angles of both swimmers are  symmetrically bounded, \ie $\bTheta \in \left[ -\thetamag,\thetamag \right]^N$. Nonetheless, enabling the flagellar swimmer to reorient, as indicated in Fig.~\ref{fig2:policy}B (\movref 2), necessitates choosing an asymmetric bound, \eg $\left[-\thetamag/2,\thetamag \right]^N$ as used here. 
This choice mimics the structural asymmetries in the flagellar scaffold of spermatozoa, which facilitates their reorientation~\cite{brokaw1979calcium,lindemann1988calcium}.

\begin{figure*}[bt!]
\centering
\includegraphics[width=1.0\linewidth]{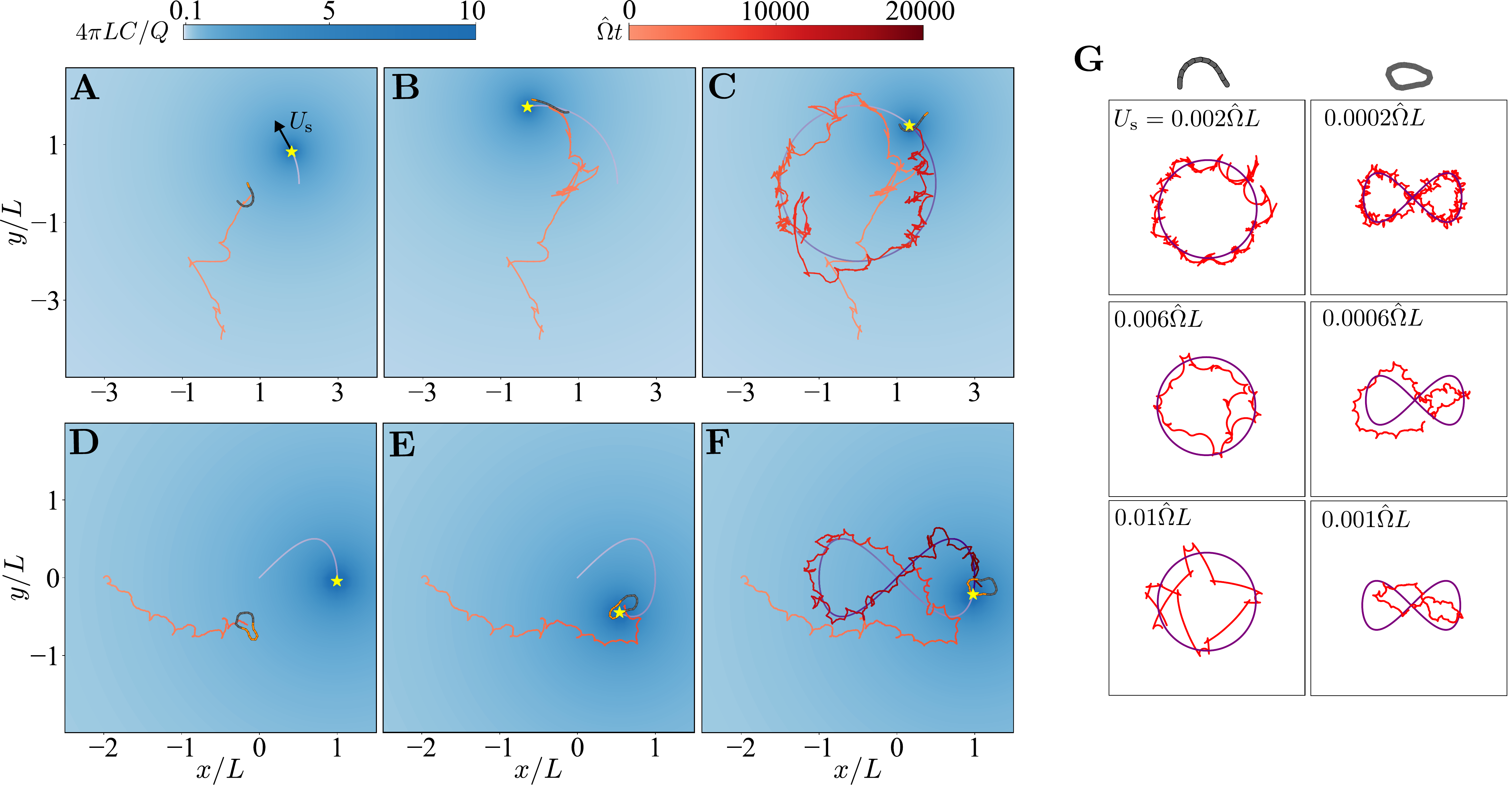}
\caption{\textbf{Chemotactic pursuit of microrobots towards a dynamic chemical source of strength $Q$.} A-C, a flagellar robot chasing (A), encountering (B), and following (C) a chemical source (symbolized by a star) moving along a circular orbit. 
D-F, similar to A-C, but for an ameboid robot and a source executing a figure-eight-shaped path.
G, trajectories of flagellar (left column) and ameboid (right column) robots pursuing a source at varying speeds.
}
\label{fig:mov}
\end{figure*}

We further demonstrate the learning curves of the two robots acquiring these primitive skills. 
As the policy evolves, the robotic agent learns to increase the front-rear pressure difference (proportional to the reward), which reaches a saturated value after $\approx 400$ policy iterations (see Fig.~\ref{fig2:policy}D). The saturation value is significantly lower for the ameboid swimmer than the flagellar counterpart. 
Realizing the linear pressure-velocity relationship in the Stokesian limit, both robotic swimmers effectively learn to propel, even though their velocities are not explicitly targeted or measured. 
This linear relationship is reflected in the translational velocities of the swimmers, which follow a similar trend to the pressure difference, as depicted in Fig.~\ref{fig2:policy}F. 
The flagellar and ameboid robots' translational velocities eventually plateau at $\Uswim \approx 0.017 \Omegamag L$ and $\Uswim \approx 0.0013 \Omegamag L$, respectively, which imply the swimming capacity of the learnt robots.
Using the same pressure-encoded reward but with the asymmetric angular bound, the flagellar swimmer develops the rotational capacity (Fig.~\ref{fig2:policy}E), therefore enabling its reorientation.

\subsubsection*{Navigation towards a static chemical source}\label{sec:static}

Based on the learnt primitive skills, the microrobotic agents are first trained to learn chemotaxis towards a stationary chemical source with strength $Q$, positioned in the swimming plane. As illustrated in Fig.~\ref{fig:nobg}A \& D (\movref 4 \& 5), both the flagellar and ameboid swimmers successfully navigate to the source located to the east of their initial positions, with the former executing a more meandering trajectory. 

To examine the resilience of the microrobotic chemotaxis in environments with multiple chemoattractants, as typically encountered in chemotaxis~\cite{jin2013gradient,petri2018neutrophil}, we introduce a disturbing chemical source with strength $\Qd$. This disturbance is positioned to the north of the swimmer's initial position and at an equal distance from it as the target source. Accordingly, this settings features two orthogonal chemoattractant gradients. In the presence of a weak disturbance $\Qd=0.2Q$, the flagellar robot consistently reaches the target source (see Fig.~\ref{fig:nobg}B). 
However, at a stronger disturbance of $\Qd=0.6Q$, the robot occasionally fails and approaches the disturbance source. At this same disturbance level, the ameboid robot shows great resilience, successfully navigating to the target in all trials, see Fig.~\ref{fig:nobg}C. Nonetheless, its success rate drops to $\approx 60\%$ when $\Qd=0.8Q$.

\begin{figure*}[bt!]
\centering
\includegraphics[width=1.0\linewidth]{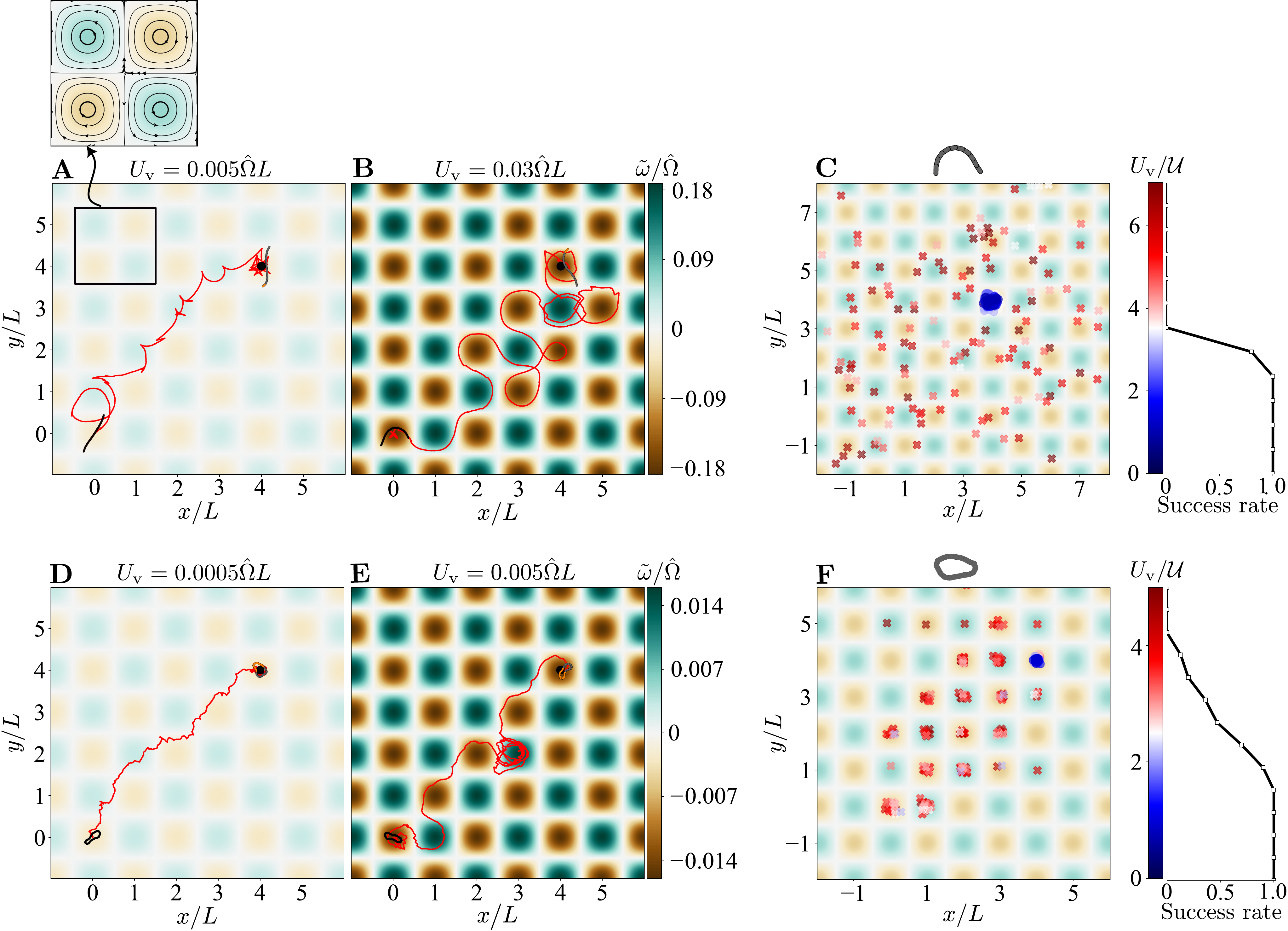}
\caption{
    \textbf{Chemotactic navigation in periodic cellular vortices with a vortical velocity $\Uv$.} Microrobots start from $(0,0)$ with the goal of reaching a chemical source positioned at $(4,4)$. A and B: a flagellar robot swims in vortices with $\Uv=0.005 \Omegamag L$ and $\Uv=0.03 \Omegamag L$, respectively. The background color denotes the dimensionless flow vorticity. 
    The inset of A depicts the typical streamlines. 
    C: statistics of successful chemotaxis versus the vortical strength $\Uv$, where crosses and circles signify failures and successes, respectively. E and F, similar to A and B, but for an ameboid swimmer, with $\Uv=0.0005 \Omegamag L$ and $\Uv=0.005 \Omegamag L$, respectively. F: similar to C, while for an ameboid swimmer.
}
\label{fig:bg}
\end{figure*}

\subsubsection*{Chasing a moving chemical source}
Having demonstrated chemotaxis towards a static chemical source, the microrobots are now tasked with pursuing a moving source that follows a predefined trajectory. This scenario emulates the behavior of white blood cells, such as neutrophils, which chase invading bacteria by tracking their chemoattractants~\cite{neutrophichase,liu2015moesin}.

The flagellar and ameboid robots are trained to purchase a chemoattractant-releasing bacterial mimic executing a circular and figure-eight path at a constant speed $\Us$, respectively. Here, $\Us = 10^{-3}\Omegamag L$ and $3\times10^{-4}\Omegamag L$ for the two swimmers. Time-lapse sequences depicted in Fig.~\ref{fig:mov} (\movref 6 \& 7) illustrate the swimmers' chemotactic navigation, highlighting their approach to the source (A, D), interception (B, E), and subsequent tracking (C, F). As shown in Fig.~\ref{fig:mov}G, both robots struggle to precisely track the source moving at an increasing speed $\Us$.

\subsubsection*{Chemotaxing within an ambient flow}
Microorganisms and microrobots typically do not swim in a quiescent environment but are commonly subject to an underlying flow. Here, we investigate how such flows affect RL-based microrobotic chemotaxis. As displayed in Fig.~\ref{fig:bg}, we navigate the articulated robots towards a static chemical source in spatially periodic cellular vortices $\tbu=\Uv \sin (\pi x/L) \cos (\pi y/L)\be_x-\Uv \cos (\pi x/L) \sin (\pi y/L)\be_y$~\cite{stommel1949trajectories}, where $\Uv$ denotes the vortical strength. Here, we have set the size of cells to match the swimmer length, $L$, serving as a representative demonstration.

In a weak cellular flow, both flagellar and ameboid robots achieves chemotaxis with ease, as indicated in Fig.~\ref{fig:bg}A \& D (\movref 8 \& 10). However, when the vortices are substantially strengthened, the robots tend to be occasionally trapped with these vortices before reaching the chemoattractant source, see Fig.~\ref{fig:bg}B and E  (\movref 9 \& 11). 
With further increased $\Uv$, the flagellar swimmer fails to reach the target, while the ameboid counterpart cannot escape from the trapping vortices (see \SIe). 
Fig.~\ref{fig:bg}C \& F characterize the success rate of robotic chemotaxis versus the vortical strength $\Uv$, with $30$ data samples collected for each strength. 
For the flagellar robot, the success rate exhibits a sharp transition from $0$ (indicating complete failure) to nearly $1$ as the ratio $\Uv/\Uswim$ decreases to approximately $3$. Conversely, for the ameboid robot, this rate gradually increases from $0$ at $\Uv/\Uswim \approx 4$ to $1$ at $\Uv/\Uswim \approx 2$, displaying an approximately linear trend.

\subsubsection*{Squeezing through a narrow constriction}
\begin{figure*}[bt!]
\centering
\includegraphics[width=1.0\linewidth]{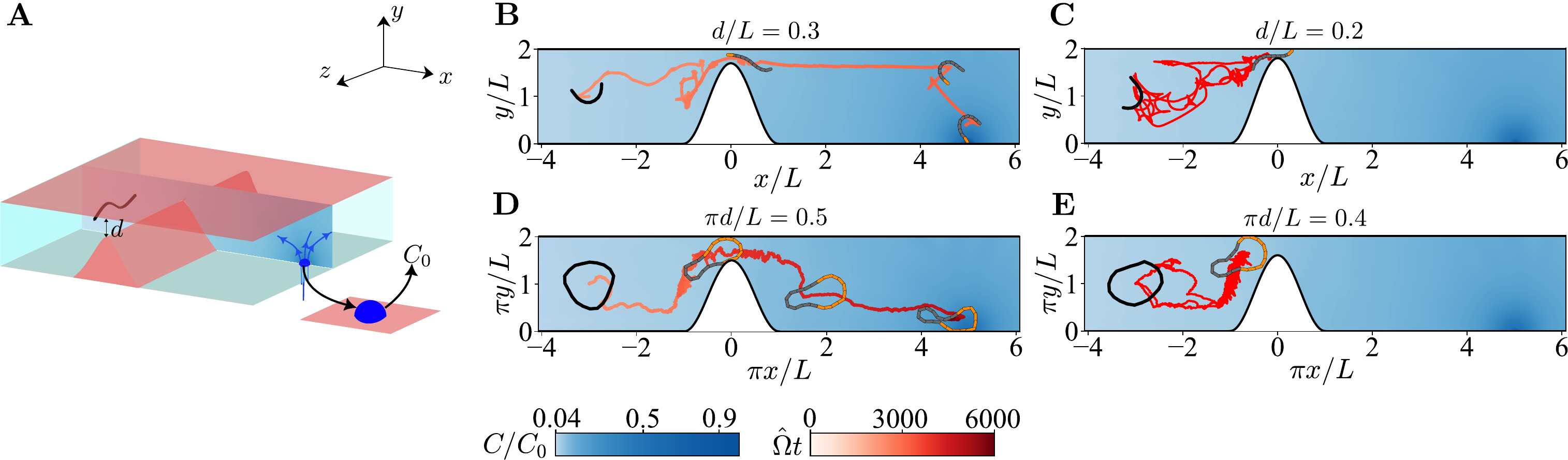}
\caption{
\textbf{Chemotactic navigation through a narrow constriction.} A, the microrobot navigates a constriction of height $d$ to reach the chemical source located on the opposite side. The source and robotic motion are within the same $xy$-plane. A flagellar robot successfully traverses a constriction of $d=0.3L$ (B), but fails to pass through a narrower one of $d=0.2L$ (C). D, similar to B, but for an ameboid robot, which squeezes through the narrow gap of $d=0.5L/\pi$, where $L/\pi$ denotes the effective diameter of ameboid robots. E, analogous to D, but with a tighter constriction of $d=0.4L/\pi$, which obstructs the swimmer. 
}
\label{fig:ob}
\end{figure*}

Microorganisms exhibit chemotaxis within complex, confined spaces. 
Human spermatozoa, measuring $\approx$ \SI{60}{\micro\meter} in length, must navigate constricted regions like cervical crypts and folds of the ampullary fallopian tube, with constrictions as tight as about \SI{100}{\micro\meter}~\cite{denissenko2012human}. More impressively, neutrophils with a typical diameter of $7\text{--}8$~\si{\micro\meter} undergo extreme cellular deformation to rapidly squeeze through $\sim$ \SI{1}{\micro\meter} tissue barriers of blood vessels during transendothelial migration---a process critical for mounting an immediate immune response~\cite{rowat2013nuclear,manley2018neutrophil}.

To emulate this biomimetic, we task our robots with navigating a 3D spanwise ($z$)-uniform channel formed by two plates (see Fig.~\ref{fig:ob}A). The bottom plate features a sinusoidal protrusion that narrows the passage to a width $d$.
The robots, initially on one side of the resulting constriction are trained to traverse the narrowed path to reach a chemical source with concentration $C_0$ located on the opposite side. 

As depicted in Fig.~\ref{fig:ob}B (\movref 12), the flagellar swimmer successfully navigates the constriction of $d/L=0.3$ towards the chemical source, yet it is obstructed by the narrower constriction of $d/L=0.2$ (see Fig.~\ref{fig:ob}C). On the other hand, the ameboid swimmer struggles more in the confined space. Fig.~\ref{fig:ob}E indicates that it cannot traverse a passage that is narrowed to $0.4$ times its effective diameter, $L/\pi$. However, by enlarging the constriction to $d = 0.5 L/\pi$, the ameboid swimmer manages to pass through, as illustrated in Fig.~\ref{fig:ob}D (\movref 13). 

We notice a pitfall in our ameboid robot; it cannot replicate the biological template---the neutrophils' capacity to squeeze through extremely tight gaps. The failure of the synthetic ameboid stems from its constant contour length $L$ that renders zero surface extensibility, in contrast to neutrophils possessing highly extensible surfaces~\cite{manley2018neutrophil}. This difference implies a potential solution leveraging elastic and flexible inter-link joints to develop soft multi-link robots.

\subsection*{Discussion}
In this study, we employ a two-level hierarchical RL to enable chemotactic navigation of an articulated microrobotic swimmer comprising sequentially connected links. The robot, via the lower-level RL, first learns the primitive skills of propulsion and reorientation. Subsequently, using these learned actions, the higher-level RL endows the robot with the ability of chemotaxis across representative biologically-inspired settings, 
including scenarios with conflicting chemoattractants, chemotactic pursuit of a moving bacterial mimic, chemotaxis in an ambient flow or through throat-like constrictions.

Depending on the topology of inter-link connectivity---linear or ring, RL imparts a topology-specific locomotory gait to the multi-link robot, mimicking the stroke of biological microswimmers with corresponding topologies. Specifically, the linearly linked robot, resembling a chain, swims by propagating a transverse bending wave characteristic of a beating flagellum; conversely, the ring-shaped counterpart undergoes longitudinal body oscillation to enable ameboid-like locomotion~\cite{barry2010dictyostelium,farutin2013amoeboid,wu2015amoeboid}. 
These two examples demonstrate that RL can instill topology-specific motility mechanisms in the articulated robot, replicating those evolved by microorganisms.

We naturally question whether such successful instillation of robotic intelligence can be extended to organisms with different topologies and respective locomotive strategies. An immediate prospect is the \textit{star} topology 
featured by octopuses and octopus-inspired robots, as depicted in \cite{fras2018bio}. We hypothesize that RL could allow multi-link robots with this topology to acquire the lesser-known arm-sculling swimming gait~\cite{kazakidi2012swimming,sfakiotakis2015octopus} of certain octopuses.
Besides imitating these natural mobility mechanisms, applying RL to the multi-link robot with topologies not found in nature could uncover novel, counterintuitive mechanisms beyond those evolved by living systems. An inspiring analogy is the RL-facilitated computer Go program, AlphaGo~\cite{silver2016mastering}, which can generate inventive winning moves unprecedented in human-to-human Go matches. Further stretching our imagination, the integration of RL with other ML techniques could empower multi-link microrobots to intelligently reconfigure their topology, adapting to environmental variations and fulfilling multitasking demands~\cite{xie2019reconfigurable}. For instance, an ameboid robot could learn to transform into a flagellar robot or to disassemble into multiple flagellar robots, and vice versa.

Beyond acquiring the motility mechanism, our microrobots have been trained to exhibit chemotaxis, in resonance with recent studies of intelligent active particles~\cite{hartl2021microswimmers,mo2023chemotaxis,nasiri2024smart,tovey2024emergence}. Besides advancing the design of medical microrobots, our research focus on programming chemotactic robotic intelligence may also inspire the development of cleaning microrobots for remediation tasks~\cite{soler2013self,parmar2018micro}, specifically targeting the removal of micro/nanoplastics, heavy metals, and radioactive contaminants from polluted waters. In addition to chemotaxis, our learning framework can be generalized to engineer diverse microrobotic taxis~\cite{you2018intelligent}, including phototaxis, rheotaxis, magnetotaxis, and combinations thereof. This generalization will promote the development of more intelligent and versatile microrobots.

Furthermore, our work has demonstrated employing hierarchical RL to achieve reset-free training of microrobots for navigation. This demonstration offers valuable insights into RL-based manipulation of microrobots in wet-lab settings---an emerging field in its infancy~\cite{muinos2021reinforcement,behrens2022smart,abbasi2024autonomous}. Given the confined and viscosity-dominated environments typical of microrobotics, eliminating manual resets of robots poses a formidable challenge for real-world microrobotic RL---a concern that remains underrecognized. Our study highlights a potential solution: leveraging hierarchical RL, and possibly other related frameworks, such as multi-task RL~\cite{kalashnikov2021mt}. This approach will facilitate  continuous RL training of microrobots with minimal human intervention, enabling them to perform dexterous medical operations under complex physiological conditions.

\newpage
\section*{Materials and Methods}
\subsection{Numerical methods}\label{sec:numerical}
We build the numerical hydrodynamic environment of swimming microrobots using a 3D boundary integral method~\cite{pozrikidis1992boundary} to solve the Stokes equation. Specifically, we adopt a regularized Stokeslet method~\cite{smith2018nearest}, with more details provided in \SIe.

\subsection{Lower-level RL}\label{sec:rll}
\subsubsection{Observations}
At the lower level of our hierarchical RL framework, the agent observes the joint angles $\bTheta$ of a microrobot as proprioceptive sensory inputs.
For ameboid swimmers, the angles are within a symmetric range, namely $\bTheta \in [-\thetamag,\thetamag]^N$. However, this situation changes for a flagellar swimmer: its propulsion admits a symmetric range, whereas its reorientation necessitates an asymmetric one, $[-\thetamag/2,\thetamag]^N$ as chosen here. Moreover, the maximum angle $\thetamag$ of flagellar swimmers is limited to
$2\pi/(N+1)$ to prevent links from crossing. For ameboid swimmers, $\thetamag = \pi/3$ is adopted to allow reasonable body deformation while avoid intersections of links. When training the robots, we judge whether an obtained action might result in limit-exceeding joint angles or crossing links. If either condition is met, a null action is implemented.

\subsubsection{Actions}
The action of an agent is encoded by a vector $\ba \in [-1,1]^N$, which correlates with the angular velocities $\bOmega$ via a linear mapping, $\bOmega = \bB \ba$. Here, the linear operator $\bB$ depends on the topology of the multi-link robot. For flagellar robots, $\bB= {\Omegamag}\bI$. In contrast, for an ameboid robot, the formulation of $\bB$ warrants special consideration due to the geometric constraints imposed by its links, which form an $N$-sided equilateral polygon.

For this polygon, the links form a closed loop, imposing the following constraints on the joint angles $\bTheta$ and uniform side length $L/N$:
\begin{align}\label{eq:con_angle}
        \sum\limits_{i=1}^N \frac{L}{N}\cos \theta_i  =\sum\limits_{i=1}^N \frac{L}{N}\sin \theta_i = 0.       
\end{align}
Taking the time derivative of Eq.~\eqref{eq:con_angle}, we obtain the constraint on $\bTheta$,
\begin{align}\label{eq:con_omega}
        \sum\limits_{i=1}^N \Omega_i \cos \theta_i =\sum\limits_{i=1}^N \Omega_i\sin \theta_i = 0,     
\end{align}
which simplifies to $\bC \bOmega = \mathbf{0}$ where
\[
\bC = \begin{bmatrix} 
    \cos \theta_1 &  \cos \theta_2 &\cdots &\cos \theta_N \\
    \sin \theta_1 &  \sin \theta_2 &\cdots &\sin \theta_N \\ 
    \end{bmatrix}.
\]
This constraint is fulfilled by setting $\bB={\Omegamag}(\bI-\bC^{+}\bC)$, with $\bC^{+}$ denoting the Moore-Penrose inverse of $\bC$.

\subsubsection{Reward}
In the lower-level RL framework, the pressure difference between the front and rear of a swimmer is used to design the reward $\reone = \lambda (\pf-\pr)$. 
For a flagellar robot, as shown in Fig.~\ref{fig1:RL}C, $\pf$ and $\pr$ are simply measured at the front and rear tips, respectively. 
The situation is slightly more complicated for an ameboid robot.
As depicted in Fig.~\ref{fig1:RL}C,  the swimmer is divided into a front half (orange links) and a rear half (grey links), 
with $\pf$ and $\pr$ the average pressures over these halves.
For an even number of links, $N$, each half contains $N/2$ links; Otherwise, the front 
and rear  halves comprise $(N-1)/2$ and $(N+1)/2$ links, respectively.

\subsubsection{Symmetry-encoded policies}\label{sec:symmetry}
The flagellar and ameboid robots possess $\nprim = 6$ and $N$ primary actions, respectively. However, exploiting the geometric symmetry of both robots can reduce the number of RL tasks for learning all these actions to two for the former and one for the latter.

Specifically, the flagellar robot requires learning policies only for forward ($\aprim=1$) and left-forward ($\aprim=2$) swimming. Policies for other primitive actions are derivable through symmetry. As illustrated in Fig.~\ref{fig1:RL}C,
for example, using left-right symmetry, the policy for right-forward movement ($\aprim=6$) can be obtained by inputting $-\bTheta$ into the policy network for left-forward swimming ($\aprim=2$). 
Likewise, fore-aft symmetry allows deriving the policy for left-backward swimming by flipping $\bTheta$ for the left-forward swimming policy ($\aprim=2$). A similar strategy applies for deriving the backward and right-backward swimming policies.

The ameboid robot features an $N$-order rotational symmetry, allowing the agent to master the policy for only one primitive action, \eg $\aprim =1$. The policy for any other action $\aprim > 1$ can be derived by permuting $\bTheta$ to $(\theta_{\aprim},...,\theta_{N},\theta_{1},...,\theta_{\aprim-1})^{\tran}$.

\subsection{Higher-level RL}\label{sec:rlh}
At the higher level, we adopt a coarse-to-refine, two-step approach. In the coarse step, we first need to identify a set of primitive actions. For the flagellar swimmer, if the concentration at the front tip exceeds that at the rear tip, the set comprises primitive actions $\{1, 2, 6\}$, which facilitate forward motion; conversely, if lower, the set is $\{3, 4, 5\}$, supporting backward motion. 
For the ameboid swimmer, the process begins by pinpointing the hinge with the highest concentration. Subsequently, any primitive action whose corresponding front half includes this hinge is included as a constituent element of the set.

\subsection{Solution of the concentration equation}\label{sec:con}
In our chemotactic navigation, the chemical sources are positioned on the $xy$-plane ($z=0$), where the microrobots are restricted to. 

As depicted in Fig.~\ref{fig:nobg}, a static chemical source of strength $Q$ is located at $\br_{\rm t}$ serving as the target, while another static source with strength $\Qd$ positioned at $\br_{\rm d}$ acts as a disturbance. The 3D chemical concentration field $C(\br)$ is governed by the equation $\grad^2 C=Q\delta^3(\br_{\rm t})+\Qd \delta^3(\br_{\rm d})$, with $\delta^3(\cdot)$ denoting the 3D Dirac delta function. 
The resultant chemical distribution can be analytically expressed as $C(\br)=Q/\left (4\pi |\br-\br_{\rm t}| \right )+\Qd/\left (4\pi |\br-\br_{\rm d}| \right)$. 
For a moving chemical source with a prescribed trajectory $\br_{\rm m}(t)$ shown in Fig.~\ref{fig:mov},
the chemical concentration is 
$C(\br,t)=Q/\left (4\pi |\br - \br_{\rm m}(t)| \right )$. 
The flagellar swimmer follows a source along the circular trajectory $\br_{\rm m}(t) =  2L \cos(\Omegamag t/1000)\be_x+ 2L \sin(\Omegamag t/1000)\be_y$; the ameboid swimmer pursues a source moving along the figure-eight path $\br_{\rm m}(t) = L \sin(\Omegamag t/4000) \be_x +  L \sin(\Omegamag t/4000) \cos(\Omegamag t/4000)\be_y$.

For navigation within a confined domain shown in Fig.~\ref{fig:ob}, the chemical field is solved numerically using COMSOL Multiphysics (I-Math, Singapore), as detailed in \SIe.

\subsection*{Description of Movies}\label{sec:movie}
Unless otherwise mentioned, the flagellar robot's number of hinges is $N=9$, while that of the ameboid robot is $N=20$.

\begin{itemize}
    \item \movref 1, a flagellar robot self-propels in the form of a transverse wave.
    \item \movref 2, a flagellar robot reorients itself.
    \item \movref 3, an ameboid robot self-propels in the form of a longitudinal wave.
    \item \movref 4, a flagellar robot swims towards a stationary chemical source.
    \item \movref 5, an ameboid robot swims towards a stationary chemical source.
    \item \movref 6, a flagellar robot chases a moving source that executes a circular path.
    \item \movref 7, an ameboid robot chases a moving source that executes a figure-eight path.
    \item \movref 8, a flagellar robot chemotaxes in weak periodic cellular vortices.
    \item \movref 9, a flagellar robot chemotaxes in strong periodic cellular vortices.
    \item \movref 10, an ameboid robot chemotaxes in weak periodic cellular vortices.
    \item \movref 11, an ameboid robot chemotaxes in strong periodic cellular vortices. 
    \item \movref 12, a flagellar robot navigates a tight constriction to reach the chemical source.
    \item \movref 13, an ameboid robot squeezes through a tight constriction to reach the chemical source.
\end{itemize}

\end{document}